\documentstyle[pasjskel,graphicx,natbib,twocolumn]{article}

\begin{document}

\title{The 1996--1997 Fading of V651 Mon, the Binary Central Star of
the Planetary Nebula NGC 2346}

\author{Taichi \textsc{Kato}}
\affil{Department of Astronomy, Faculty of Science, Kyoto University,
       Sakyo-ku, Kyoto 606-8502}
\email{tkato@kusastro.kyoto-u.ac.jp}

\author{Daisaku \textsc{Nogami}}
\affil{Hida Observatory, Kyoto University, Gifu 506-1314}
\email{nogami@kwasan.kyoto-u.ac.jp}

\email{\rm{and}}

\author{Hajime \textsc{Baba}}
\affil{Astronomical Data Analysis Center, National Astronomical
       Observatory, Mitaka, Tokyo 181-8588}
\email{hajime.baba@nao.ac.jp}


\begin{abstract}
   V651 Mon is the binary central star of the bipolar planetary nebula
NGC 2346.  The star showed the second-ever deep fading in 1996--1997,
which was presumably caused by obscuration by a dust cloud in the
planetary nebula, as was proposed to explain the 1981--1985 event.
The entire duration of the 1996--1997 event was $\sim$400 d,
remarkably shorter than the 1981--1985 event, suggesting that the
obscuring body was smaller or had a larger tangential velocity.
The most remarkable feature in this event was the presence of a sharply
defined transient clearing (brightening).  From the time-scale of the
variation, we propose an upper limit of the projected scale of several
times $\sim$10$^{11}$ cm of the structure responsible for the brightening.
This observation provides the first evidence for a sharply defined,
small lucent structure within the obscuring body around the central star
of NGC 2346
\end{abstract}


\section{Introduction}

   V651 Mon is the binary central star of the bipolar planetary nebula
NGC 2346.  The 15.991-d binary consists of an A5 V star, which emits most
of the visual light, and a hot ($T_{\rm eff}$ $\sim$10$^5$ K) subdwarf,
which emits the most of the ultraviolet light \citep{men81}.  This binary
has received special attention since its fading episode starting in 1981
\citep{koh82}.

   The fading initially showed a strong modulation with the orbital period.
Subsequent observations led to a picture of a passing cloud (within the
planetary nebula) in front of the binary: eclipse-like fadings in the
visual light occur when the orbiting A5 V star passes behind the cloud
(for a review on this object, see \cite{cos86}).
\citet{cos86} expect, from the amount of dust from IRAS observations,
that the probability of such passages is very small.
In fact, plate searches could not detect any similar events between 1899
and 1981 \citep{sch83}.  Further detailed photoelectric observations
only caught small variations \citep{koh92}, or a transient short fading
\citep{gro93}.

   In spite of a prediction by \citet{cos86}, the second-ever deep
fading was reported in 1996 September \citep{ove96}.  We started CCD
observations upon this alert.

\section{Observation}

   The observations were made on 38 nights between 1996 November 14 and
1998 January 2, using a CCD camera (Thomson TH~7882, 576 $\times$ 384 pixels,
on-chip 2 $\times$ 2 binning adopted) attached to the Cassegrain focus of
the 60 cm reflector (focal length=4.8 m) at Ouda Station, Kyoto University
\citep{oht92}.  An interference filter was used, which had been
designed to reproduce the Johnson {\it V} band.  The exposure time was
20--60 s, depending on the brightness of the object.  The frames were
first corrected for standard de-biasing and flat fielding, and were then
processed by a microcomputer-based photometry package developed by one
of the authors (TK).  We employed a PSF profile fitting procedure for
the central star in order to minimize any contamination from the
planetary nebula.
The magnitudes were determined relative to GSC 4815.3644 ($V$=11.76,
VSNET chart), whose constancy during the run was confirmed using
GSC 4815.3177.  Table \ref{tab:log} lists the log of observations.

\begin{table}
\caption{Log of observations}\label{tab:log}
\begin{center}
\begin{tabular}{cccc}
\hline\hline
Mid-JD$^*$ & Mean mag$^\dagger$ & Error$^\ddagger$ & N$^\|$ \\
\hline
50402.322 &  1.195   & 0.038 &  5 \\
50404.199 &  0.685   & 0.054 &  3 \\
50407.247 &  1.120   & 0.043 &  5 \\
50427.089 &  1.860   & 0.025 &  5 \\
50429.107 &  1.537   & 0.021 &  5 \\
50432.160 &  1.854   & 0.021 &  5 \\
50438.142 &  1.626   & 0.056 &  3 \\
50439.044 &  1.478   & 0.088 &  3 \\
50441.050 &  2.142   & 0.030 &  3 \\
50442.050 &  1.935   & 0.039 &  3 \\
50443.076 &  1.540   & 0.051 &  3 \\
50445.094 &  1.164   & 0.016 &  5 \\
50448.078 &  1.582   & 0.029 &  5 \\
50449.016 &  2.095   & 0.065 &  3 \\
50450.128 &  2.010   & 0.160 &  2 \\
50451.033 &  1.911   & 0.083 &  6 \\
50452.238 &  1.805   & 0.021 &  3 \\
50452.969 &  1.410   & 0.014 &  3 \\
50455.110 &  1.925   & 0.015 &  5 \\
50461.094 &  1.611   & 0.062 &  3 \\
50462.073 &  1.538   & 0.024 &  3 \\
50464.107 &  2.098   & 0.036 &  3 \\
50468.998 &  2.133   & 0.055 &  3 \\
50507.048 &  2.496   & 0.015 &  3 \\
50507.960 &  2.279   & 0.025 &  5 \\
50509.037 &  2.296   & 0.021 &  5 \\
50512.966 &  2.495   & 0.017 &  5 \\
50515.988 &  2.401   & 0.037 &  5 \\
50518.948 &  2.090   & 0.036 &  5 \\
50535.917 &  2.249   & 0.095 &  7 \\
50537.921 &  1.500   & 0.012 &  5 \\
50538.930 &  1.162   & 0.028 &  3 \\
50539.921 &  0.479   & 0.089 &  3 \\
50551.924 &  2.236   & 0.019 &  5 \\
50552.938 &  2.204   & 0.058 &  5 \\
50711.326 & $-$0.180 & 0.022 &  5 \\
50814.133 & $-$0.442 & 0.053 &  3 \\
50816.157 & $-$0.581 & 0.027 &  3 \\
\hline
 \multicolumn{4}{l}{$^*$ JD$-$2400000.} \\
 \multicolumn{4}{l}{$^\dagger$ Magnitude relative to GSC 4815.3644.} \\
 \multicolumn{4}{l}{$^\ddagger$ Standard error of nightly average.} \\
 \multicolumn{4}{l}{$^\|$ Number of frames.} \\
\end{tabular}
\end{center}
\end{table}

\section{Discussion}

\begin{figure*}
  \begin{center}
    \FigureFile(130mm,90mm){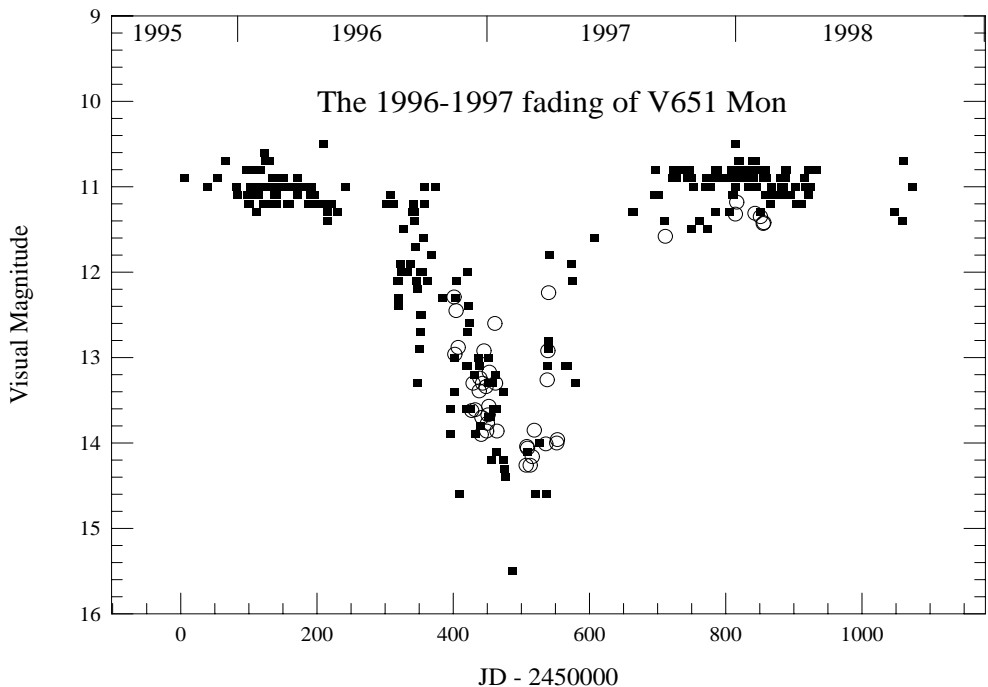}
  \end{center}
  \caption{Visual light curve of the 1996--1997 fading from observations
  reported to VSNET (filled squares).  CCD $V$-band observations (open
  circles) from table \ref{tab:log} are also shown.
  The total duration of the deep fading episode was $\sim$400 d.}
  \label{fig:fading}
\end{figure*}

   Figure \ref{fig:fading} presents the overall course of the 1996--1997
fading episode.  Visual observations were reported to
VSNET\footnote{$\langle$ http://www.kusastro.kyoto-u.ac.jp/vsnet/ $\rangle$.}
(VSNET is an international network of variable star observers,
designed to make collaborative studies on selected variable stars).
All observers used comparison stars calibrated in the $V$-band.
The present CCD $V$-band observations are also shown (open circles).
The total duration of the event was $\sim$400 d,
which is remarkably shorter than the 1981--1985 event.

\begin{figure}
  \begin{center}
    \FigureFile(88mm,60mm){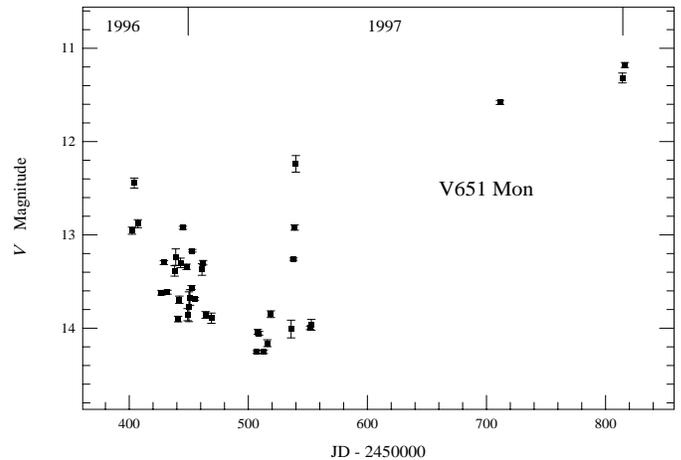}
  \end{center}
  \caption{Light curve of V651 Mon from the present CCD observations.}
  \label{fig:lc}
\end{figure}

   Figure \ref{fig:lc} shows an overall light curve from our CCD
observations.  The CCD observation started when the 1996--1997 fading
episode reached close to its minimum.  The object recovered to its
normal magnitude during the latest two observations (at the end of 1997).

   Following the interpretation by \citet{cos86}, the cloud responsible
for the fade was either smaller or had a larger tangential velocity.
The total depth of the fading ($\sim$2.7 mag) may be slightly smaller
than the extreme value reported in the 1981--1985 event.  This difference,
however, may be a result of a systematic difference of the measurements:
in the 1981--1985 event, the aperture background subtraction technique
in photoelectric photometry was employed in order to measure the faint
object against the bright nebular background, while the present observations
of the 1996--1997 event used PSF fitting of the stellar profile on
two-dimensional CCD images.
By taking this effect into account, the depth was found to be roughly
comparable to that of the 1981--1985 event, implying that the opacity
of the cloud is similar.

\begin{figure}
  \begin{center}
    \FigureFile(88mm,60mm){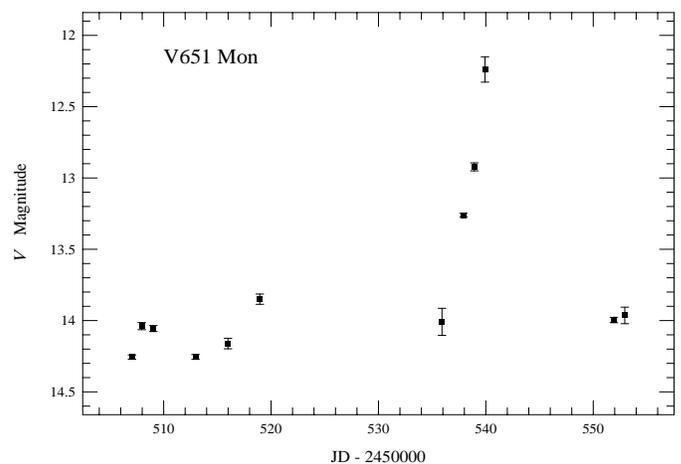}
  \end{center}
  \caption{Transient, sharp brightening observed during the fading
  episode.}
  \label{fig:brightening}
\end{figure}

   Another notable feature of the present event is the presence of a
sharply defined transient brightening around JD 2450540.  The enlarged
light curve around this brightening is shown in figure
\ref{fig:brightening}.
Although visual observations reported to VSNET suggest the presence of
a weak 16-d periodicity, as was observed in the 1981--1985 event,
no other brightening with a similar amplitude was observed around the
binary phase of the brightening.  Since the binary motion of the central
close binary is expected to be much faster than the tangential
velocity of the orbiting cloud [\citet{cos86} proposed 0.14 km s$^{-1}$
for the 1981--1985 event], the sharply defined brightening is considered
to more reflect the orbital motion of the A5 V star against the
foreground cloud.  By assuming $K_1$=16.4 km s$^{-1}$ \citep{men81},
the sharp structure of the light curve corresponds to an upper limit of
the projected scale of several times $\sim$10$^{11}$ cm.  This scale is
by a factor of about ten smaller than the entire cloud responsible for
the 1981--1985 fade, which was estimated to have a size of 2--5 $\times$
10$^{12}$ cm \citep{cos86}.

   This observation not only provides the first evidence for such a sharply
defined, small lucent structure within the obscuring body around the
central star of NGC 2346, but also the first-ever direct determination
of the dimension of a fine structure within a dust cloud of a planetary
nebula.

\vskip 3mm

   The authors are grateful to VSNET members for providing vital
observations.

\end{document}